\documentclass[aps,floats,floatfix,showpacs,preprintnumbers,amssymb,prl,twocolumn,superscriptaddress,nofootinbib,nolongbibliography,reprint]{revtex4-2}

\usepackage[T1]{fontenc}
\usepackage[utf8]{inputenc}               
\usepackage{hyperref}
\hypersetup{
    colorlinks=true,
    linkcolor=blue,
    citecolor=blue,
    urlcolor=blue
}

\usepackage{amssymb,amsmath,verbatim,mathtools,needspace,enumitem,etoolbox,graphicx,microtype,afterpage,xspace,tabularx,lmodern,multirow, makecell, booktabs}

\makeatletter
\providecommand\MT@suspend@tagging{}
\makeatother
\DeclareMathOperator{\Tr}{Tr}
\usepackage{appendix}
\usepackage{tensor}
\usepackage{array}
\usepackage[normalem]{ulem}
\usepackage[dvipsnames, usenames]{xcolor}
\usepackage{xr-hyper}
\definecolor{linkcolor}{rgb}{0.0,0.3,0.5}
\usepackage[all]{hypcap}
\usepackage[T1]{fontenc}
\usepackage[utf8]{inputenc}
\usepackage[usenames,dvipsnames]{xcolor}
\definecolor{cornellGreen}{HTML}{6EB43F}
\definecolor{cornellRed}{HTML}{B31B1B}
\hypersetup{colorlinks=true,citecolor=cornellRed,linkcolor=cornellRed,urlcolor=cornellRed}
\usepackage{multirow}

\definecolor{romared}{RGB}{142,0,28}

\def\bb#1\ee{\begin{equation}\begin{split}#1\end{split}\end{equation}}

\def\({\left(}
\def\){\right)}
\def\[{\left[}
\def\]{\right]}

\def\be{\begin{equation}}
\def\ee{\end{equation}}
\newcommand{\beq}{\begin{eqnarray}}
\newcommand{\eeq}{\end{eqnarray}}

\usepackage{makecell}

\newcolumntype{Y}{>{\centering\arraybackslash}X}

\usepackage{graphicx}
\usepackage{dcolumn}
\usepackage{bm}
\usepackage{hyperref}
\usepackage{color}

\usepackage{times} 

\usepackage{graphicx} 
\usepackage{booktabs} 

\usepackage{amsfonts, amsmath, amsthm, amssymb} 

 \definecolor{mypurple}{RGB}{130, 0, 130} 

\begin{document}
\affiliation{Institute of Fundamental Physics and Quantum Technology, \& School of Physical Science and Technology, Ningbo University, Ningbo, Zhejiang 315211, China}

\title{Exact stress-tensor generators distinguish integrable continuations}

\author{H. Babaei-Aghbolagh}
\email{hosseinbabaei@nbu.edu.cn}
\affiliation{Institute of Fundamental Physics and Quantum Technology, \& School of Physical Science and Technology, Ningbo University, Ningbo, Zhejiang 315211, China}
\author{Song He }
\email{ hesong@nbu.edu.cn}
\affiliation{Institute of Fundamental Physics and Quantum Technology, \& School of Physical Science and Technology, Ningbo University, Ningbo, Zhejiang 315211, China}
\affiliation{Max Planck Institute for Gravitational Physics (Albert Einstein Institute), Am M\"uhlenberg 1, 14476 Golm, Germany}

\author{Zahra Pezhman}
\email{zpezhmanh@phd.guilan.ac.ir}
\affiliation{Department of Physics, University of Guilan, P.O. Box 41335-1914, Rasht, Iran}

\date{\today}
\begin{abstract}
The same leading $T\bar T$ interaction can coexist with sharply different strong-field responses, even within causal, classically integrable sigma models. We reconstruct exact autonomous stress-tensor generators for prescribed one-scale Courant--Hilbert families in two dimensions. Separating the invariant stress scale from a signed trace ratio reduces reconstruction to one scalar inversion on each invertible physical branch. The resulting profile determines both the deformation law and the relation between local energy density and pressure. For generalized Born--Infeld and logarithmic models, the leading interactions coincide, yet pressure saturates in the former and grows without bound in the latter. Their exact generators are respectively quadratic and nonpolynomial, the latter involving Lambert $W_{-1}$. A regular Legendre transformation gives the same profile in the auxiliary potential formulation. The flows commute locally with root-$T\bar T$ within a common smooth branch and retain a common classical Lax form. Exact reconstruction thus quantifies the nonlinear freedom left by leading-order matching.

\end{abstract}

\maketitle

\section{Introduction}
The $T\bar T$ deformation demonstrates that an irrelevant interaction can remain exactly tractable at finite coupling~\cite{Zamolodchikov:2004ce,Smirnov:2016lqw,Cavaglia:2016oda,Bonelli:2018kik} (reviews in~\cite{Jiang:2019epa,He:2025ppz}). Its defining law is quadratic in the stress tensor and preserves integrability~\cite{Conti:2018jho,Chen:2021aid}, with a random-geometry interpretation~\cite{Cardy:2018sdv}. Auxiliary field constructions enlarge this setting to an infinite family of classically integrable sigma models~\cite{Ferko:2024ali,Bielli:2024ach,Bielli:2024khq}. Within this family, two causal models with the same leading $T\bar T$ interaction can have qualitatively different pressure behavior at large energy density: saturation or unbounded growth. This illustrates the known freedom in higher terms of the generating function~\cite{Russo:2024xnh}, now expressed through invariant stress responses. Sharing the leading interaction does not mean satisfying the same full $T\bar T$ flow equation.

To quantify this difference, we ask which exact stress-tensor operator generates a prescribed integrable Lagrangian family and how it encodes the strong-field response. The Courant--Hilbert (CH) construction~\cite{Courant:Hilbert}, used to describe four-dimensional causal self-dual electrodynamics~\cite{Russo:2024llm,Russo:2024xnh}, parametrizes the relevant sigma models by one function~\cite{Babaei-Aghbolagh:2025uoz,Babaei-Aghbolagh:2025hlm,Baglioni:2025tsc}. Exact local reconstruction by stress-tensor inversion is already established in self-dual electrodynamics~\cite{Ferko:2023wyi}. General stress-tensor flows and autonomous invariant expansions also describe these deformations~\cite{Ferko:2024ali,Babaei-Aghbolagh:2024uqp,Babaei-Aghbolagh:2025hlm}, including one-scale CH families matched through sixth order in the coupling in four-dimensional electrodynamics~\cite{Babaei-Aghbolagh:2026vkm}. 
In four-dimensional electrodynamics, Ref.~\cite{pezhman2026exacttbartlikeflowsnonlinear} applied the reconstruction to prescribed one-scale Courant-Hilbert families. There, a dimensionless signed stress ratio eliminates the explicit coupling dependence, gives exact profiles without order-by-order matching, and makes the single-valued physical branches manifest.

We consider classical two-parameter families connected to the principal chiral model (PCM), with one dimensionful coupling $\lambda$ and a marginal root-$T\bar T$ coupling $\gamma$~\cite{Ferko:2022cix,Babaei-Aghbolagh:2022leo,Borsato:2022tmu,Babaei-Aghbolagh:2024hti}. Separating the stress-tensor scale from a dimensionless signed trace ratio yields
\begin{equation*}
\partial_\lambda\mathcal L=\frac{2X-Y}{4}\,\Psi(r),
\qquad
r=-\frac{T^\mu{}_\mu}{\sqrt{2X-Y}},
\end{equation*}
where $X=T_{\mu\nu}T^{\mu\nu}$ and $Y=(T^\mu{}_\mu)^2$. The ratio $r$ measures the departure from a traceless classical stress tensor. The model-dependent profile $\Psi$ follows by inverting a single scalar relation fixed by the CH generating function. On each physical branch where this relation is invertible, the flow contains no explicit coupling or current dependence. It thus gives an autonomous deformation law for the complete trajectory on that branch.

The generalized Born--Infeld and logarithmic models turn the distinction into an invariant comparison at fixed energy density and the same positive coupling. The reconstructed profiles yield exact relations between the stress-tensor eigenvalues, with Born--Infeld pressure strictly below logarithmic pressure throughout the positive branch. The former tends to $1/\lambda$; the latter grows logarithmically without bound. On regular Legendre patches, the same profile follows from the potential in the equivalent Russo--Townsend auxiliary field formulation~\cite{Russo:2025fuc,Baglioni:2025tsc}. Within a common smooth physical branch, the reconstructed flow commutes locally with root-$T\bar T$ and the family retains a common classical Lax form. The exact profile therefore resolves a strong-field distinction that the shared leading interaction cannot determine.

\section{Exact reconstruction of the stress-tensor flow}\label{sec:CH}
We reconstruct the deformation generator by separating the stress-tensor scale from a dimensionless coordinate along the prescribed trajectory. Consider a group sigma model with Maurer--Cartan currents $j_\pm$ and invariants $P_1=-\Tr[j_+j_-]$ and $P_2=\tfrac12\big(\Tr[j_+j_+]\Tr[j_-j_-]+P_1^2\big)$. In the variables $u,v=\tfrac14(\sqrt{2P_2-P_1^2}\mp P_1)\ge0$, the condition $\mathcal L_u\mathcal L_v=-1$ yields a classical Lax representation of PCM type~\cite{Faddeev:1987ph,Babelon:2003qt,Ferko:2024ali,Bielli:2024ach}. Its Courant--Hilbert solution is
\begin{equation}
\mathcal L=\ell(\tau)-\frac{2u}{\ell'(\tau)},\qquad
\tau=v+\frac{u}{\ell'(\tau)^2}.
\label{CH}
\end{equation}
We restrict to causal branches with $\ell'\ge1$ and $\ell''\ge0$~\cite{Russo:2024xnh}. The implicit constraint then defines $\tau$ locally and smoothly, since $1+2u\ell''/\ell'^3>0$.

Fix a dimensionless generating function $f$, normalized by $f(0)=0$ and $f'(0)=1$, and consider the one-scale family
\begin{equation}
\begin{aligned}
\ell&=\lambda^{-1}f(\zeta),&\quad \zeta&=e^\gamma\lambda\tau,\\
a&=\zeta f'(\zeta),& b&=f(\zeta)-\zeta f'(\zeta).
\end{aligned}
\label{onescale}
\end{equation}
Here $[\lambda]=-2$, while $\gamma$ is dimensionless. At fixed $\gamma$, the continuous $\lambda\to0$ limit is $\mathcal L_0=e^\gamma v-e^{-\gamma}u$, the two-dimensional ModMax model; the PCM is recovered at $\gamma=0$. At fixed $u,v$, differentiating the constraint in Eq.~\eqref{CH} cancels the implicit $\tau$ dependence and gives $\partial_\alpha\mathcal L|_{u,v}=\partial_\alpha\ell|_\tau$ for either coupling. Hence
\begin{equation}
\partial_\gamma\mathcal L=\tau\ell'=\frac a\lambda,
\qquad
\partial_\lambda\mathcal L=-\frac b{\lambda^2}.
\label{lemma}
\end{equation}
The remaining task is to express these derivatives through the stress tensor alone.

The stress tensor takes the form $T_{\mu\nu}=(a/\lambda)S_{\mu\nu}+(b/\lambda)g_{\mu\nu}$, where $S^\mu{}_\mu=0$ and $S_{\mu\nu}S^{\mu\nu}=2$. Its eigenvalues are $p=\ell$ and $-\mathcal E=\ell-2\tau\ell'$. Thus $T^\mu{}_\mu=2b/\lambda$, and the quadratic invariants satisfy
\begin{equation}
X=\frac{2(a^2+b^2)}{\lambda^2},\qquad
Y=\frac{4b^2}{\lambda^2}.
\label{XY}
\end{equation}
The rescaled invariants $\lambda^2X$ and $\lambda^2Y$ depend only on $\zeta$. On a physical branch with $\tau>0$, $a/\lambda=\tau\ell'>0$, so their common scale is removed by
\begin{equation}
r=-\frac ba=-\frac{T^\mu{}_\mu}{\sqrt{2X-Y}}
=\frac{\mathcal E-p}{\mathcal E+p}.
\label{rdef}
\end{equation}
Whenever $r(\zeta)$ is locally invertible on a chosen interval, define $\Psi(r)=r/a(\zeta(r))$. Eliminating the scale between Eqs.~\eqref{lemma} and \eqref{XY} gives the exact generators
\begin{equation}
\boxed{\begin{aligned}
\partial_\lambda\mathcal L&=\frac{2X-Y}{4}\,\Psi(r),\\
\partial_\gamma\mathcal L&=\frac12\sqrt{2X-Y}.
\end{aligned}}
\label{main}
\end{equation}
The prescribed function $f$ fixes the irrelevant profile, whereas the root-$T\bar T$ generator is universal. Both flows are autonomous: their generators have no explicit dependence on either coupling or the currents. Invertibility is an additional condition, not a consequence of causality; turning points require separate branches. At zero invariant scale, expressions involving $r$ are understood through the continuous limit of the chosen family. The linear family $f=\zeta$ has a vanishing irrelevant flow and requires no inversion.

The weak-field expansion connects this reconstruction to the perturbative description. For analytic $f=\zeta+a_1\zeta^2+a_2\zeta^3+a_3\zeta^4+\cdots$ with $a_1\ne0$, use $\rho=\sqrt{Y/X}$ on the causal branch $r\ge0$, where $\rho^2=2r^2/(1+r^2)$.

\section{Same leading interaction, distinct strong-field responses}\label{sec:examples}\label{sec:causal}
The generalized Born--Infeld and logarithmic families provide a direct test of what the exact reconstruction distinguishes. Their CH generating functions are~\cite{Russo:2024llm}
\begin{equation}
f_{\rm BI}(\zeta)=1-\sqrt{1-2\zeta},\qquad
f_{\log}(\zeta)=-\ln(1-\zeta).
\label{comparisonf}
\end{equation}

For both models, the causal branches connected to the weak-field seed have $\lambda>0$ and $\gamma\ge0$, with $0\le\zeta<1/2$ for Born--Infeld and $0\le\zeta<1$ for the logarithmic family. On these intervals $\ell'\ge1$ and $\ell''\ge0$. Convexity and $\ell(0)=0$ imply~\cite{Russo:2024xnh}
\begin{equation}
\mathcal E-p=2(\tau\ell'-\ell)=-\frac{2b}{\lambda}\ge0.
\label{causalE}
\end{equation}
For $\tau>0$, $p=\ell\ge\tau>0$, hence
\begin{equation}
0\le r<1,\qquad 0\le\rho<1.
\label{window}
\end{equation}
The inequalities constrain the physical domain; the required invertibility is checked separately. For these two families, $r(\zeta)$ is strictly increasing throughout the stated intervals.

For Born--Infeld, $r=(1-\sqrt{1-2\zeta})/(1+\sqrt{1-2\zeta})$. Inverting this relation in Eq.~\eqref{main} gives
\begin{equation}
\Psi_{\rm BI}(r)=\frac{1-r^2}{2},\qquad
\partial_\lambda\mathcal L_{\rm BI}=\frac{X-Y}{4}.
\label{flowBI}
\end{equation}
The reconstruction thus recovers the quadratic $T\bar T$ generator of the generalized Born--Infeld model~\cite{Conti:2018jho}, with no higher invariant terms.

For the logarithmic family, set $\hat y=(1-\zeta)^{-1}>1$. The trace ratio becomes $r=1-\ln\hat y/(\hat y-1)$, and its inversion yields
\begin{equation}
\Psi_{\log}(r)=\frac{r(r-1)}{W_{-1}\big((r-1)e^{r-1}\big)-(r-1)}.
\label{psilog}
\end{equation}
The lower real branch $W_{-1}$ of the Lambert function~\cite{Corless:1996} selects $\hat y>1$ for $0<r<1$. At $r=0$, the expression is defined by continuity, giving $\Psi_{\log}(0)=1/2$. Equation~\eqref{psilog} specifies the exact nonpolynomial generator on this entire branch.

The distinction is visible directly in the profiles of $\partial_\lambda\mathcal L=X\mathcal F(\rho)$:
\begin{equation}
\begin{aligned}
\mathcal F_{\rm BI}(\rho)&=\tfrac14(1-\rho^2),\\
\mathcal F_{\log}(\rho)&=\tfrac14-\tfrac{\sqrt2}{6}\rho-\tfrac7{72}\rho^2+O(\rho^3).
\end{aligned}
\label{comparisonF}
\end{equation}
Both have $C_0=1/4$, but the logarithmic flow has $C_1=-\sqrt2/6$. Its first correction is proportional to $\sqrt{XY}$, already excluding a polynomial in $X,Y$, although the flow is analytic in $\zeta$ near zero. 

The profile also determines a physical comparison without referring to the auxiliary coordinate $\tau$. Since $\lambda\mathcal E=a(1+r)$, $\lambda p=a(1-r)$, and $\Psi=r/a$, it gives
\begin{equation}
\lambda\mathcal E=\frac{r(1+r)}{\Psi(r)},\qquad
\lambda p=\frac{r(1-r)}{\Psi(r)}.
\label{profilestress}
\end{equation}
This parametrization applies where $\Psi\ne0$, with degenerate endpoints treated by continuity. Eliminating $r$ for the two models yields
\begin{equation}
\begin{aligned}
p_{\rm BI}&=\frac{\mathcal E}{1+\lambda\mathcal E},\\
\lambda\mathcal E&=2\bigl(e^{\lambda p_{\log}}-1\bigr)-\lambda p_{\log}.
\end{aligned}
\label{stressresponse}
\end{equation}
These are relations between the local Hilbert stress-tensor eigenvalues $(-\mathcal E,p)$, independent of $\gamma$ on the stated branches; no thermal ensemble is assumed. The logarithmic relation is strictly increasing in $p\ge0$ and therefore single-valued. At fixed $\mathcal E>0$ and the same $\lambda>0$,
\begin{equation}
0<p_{\rm BI}(\mathcal E)<p_{\log}(\mathcal E)<\mathcal E.
\label{pressureorder}
\end{equation}
To establish the strict comparison, write $t=\lambda p$. For $0<t<1$, the inverse curves obey $\lambda\mathcal E_{\rm BI}=t/(1-t)>2(e^t-1)-t=\lambda\mathcal E_{\log}$, as their Taylor coefficients first differ at $t^3$ and every subsequent coefficient is larger for Born--Infeld. Both curves are strictly increasing, giving Eq.~\eqref{pressureorder} on the entire positive-energy branch. Near zero, $p_i=\mathcal E-\lambda\mathcal E^2+c_i\lambda^2\mathcal E^3+O(\lambda^3\mathcal E^4)$, with $c_{\rm BI}=1$ and $c_{\log}=5/3$. With the stress normalization fixed, the quadratic term fixes any field-independent redefinition of $\lambda$; the unequal cubic terms then preclude matching the two stress curves by that redefinition.


The strong-field separation is qualitative: as $\mathcal E\to\infty$, Eq.~\eqref{stressresponse} gives $p_{\rm BI}\to1/\lambda$, whereas $p_{\log}\sim\lambda^{-1}\ln(\lambda\mathcal E/2)$ grows without bound.
The corresponding CH endpoints are $\tau_c=e^{-\gamma}/(2\lambda)$ and $e^{-\gamma}/\lambda$. In both models, $p/\mathcal E\to0$ and $r,\rho\to1$. These bounds concern $\tau$; the allowed $u$ direction remains unbounded. Both normalized profiles vanish, but the growing stress scale makes the full irrelevant generators diverge. 

\section{Equivalent potential and compatible classical flows}\label{sec:RT}\label{sec:comm}
The reconstruction can also be read from the scalar potential in the Russo--Townsend formulation~\cite{Russo:2025fuc,Kuzenko:2025gvn}, applied to two-dimensional sigma models in Refs.~\cite{Babaei-Aghbolagh:2025uoz,Baglioni:2025tsc}. On an interval with $\ell''>0$, the Legendre transform $\Omega(y)=y\tau-\ell(\tau)$ gives $y=\ell'$ and $\Omega_y=\tau$. The equivalent auxiliary Lagrangian is
\begin{equation}
\mathcal L(u,v;y)=yv-\frac{u}{y}-\Omega(y).
\label{LRT}
\end{equation}
Its stationary condition, $\Omega_y=v+u/y^2$, reproduces the CH constraint. Linear seeds are degenerate limits with fixed $y=e^\gamma$, including the PCM at $\gamma=0$; $y$ cannot then be varied as an unconstrained auxiliary field.

The potential separates the trace from the remaining stress-tensor scale. Using $\tau\ell'=y\Omega_y$ and $\ell-\tau\ell'=-\Omega$, one obtains
\begin{equation}
\begin{aligned}
T^\mu{}_\mu&=-2\Omega,\qquad Y=4\Omega^2,\\
X&=2\big[(y\Omega_y)^2+\Omega^2\big].
\end{aligned}
\label{XYOmega}
\end{equation}
Thus $\Omega$ carries the trace information that enters $r$, while $y\Omega_y$ supplies the positive scale $\sqrt{2X-Y}/2$ on the branch with $\tau>0$.

For the one-scale family, write $\Omega=\lambda^{-1}\omega(\hat y)$, where $\hat y=e^{-\gamma}y=f'(\zeta)$ and $\omega=\hat y\zeta-f(\zeta)$. Stationarity gives $\partial_\alpha\mathcal L|_{u,v}=-\partial_\alpha\Omega|_y$ for either coupling, hence
\begin{equation}
\begin{aligned}
\partial_\lambda\mathcal L&=\frac{\Omega}{\lambda},&
\partial_\gamma\mathcal L&=y\Omega_y,\\
r&=\frac{\omega}{\hat y\omega'},&
\Psi&=\frac{\omega}{(\hat y\omega')^2}.
\end{aligned}
\label{flowsRT}
\end{equation}
Here $\omega'=d\omega/d\hat y$. Since $\omega=-b$ and $\hat y\omega'=a$, eliminating $\hat y$ gives exactly the profile in Eq.~\eqref{main}, subject to the same branch-inversion condition. 

The two reconstructed flows are compatible within the common smooth family. Differentiating the CH constraint at fixed $u,v$ gives the exact identity
\begin{equation}
\partial_\alpha\partial_\beta\mathcal L
=\ell_{\alpha\beta}-\frac{2u\,\ell'_\alpha\ell'_\beta}{\ell'^3+2u\ell''},
\qquad \alpha,\beta\in\{\lambda,\gamma\},
\label{mixedflow}
\end{equation}
where coupling derivatives of $\ell$ are taken at fixed $\tau$. The denominator is positive on the causal branch, and the expression is symmetric in $\alpha,\beta$. Consequently, the irrelevant and root-$T\bar T$ deformations commute locally on each $C^2$ branch; finite changes can be interchanged when both paths remain on that same branch. This analytic statement is consistent with perturbative compatibility checks in four-dimensional self-dual theories~\cite{Babaei-Aghbolagh:2024uqp}.

Classical integrability is preserved because the trajectories remain in the CH class. Its equations of motion admit the PCM-type Lax form~\cite{Ferko:2024ali,Bielli:2024ach,Baglioni:2025tsc}
\begin{equation}
\mathfrak L_\pm(z)=\frac{j_\pm\pm z\mathfrak J_\pm}{1-z^2},
\label{Laxflow}
\end{equation}
where $z$ is the spectral parameter and $\mathfrak J_\pm$ are the model's conserved Noether currents, reducing to $j_\pm$ in the PCM. Their dependence on the physical fields changes with the potential, while the rational Lax form is common. In the corresponding regular auxiliary-field description, the established Hamiltonian analysis gives the PCM Maillet structure~\cite{Maillet:1985ek,Maillet:1985ec}, with an interaction-independent twist function in fixed canonical conventions~\cite{Ferko:2024ali,Bielli:2024ach,Baglioni:2025tsc}. With periodic spatial boundary conditions, this supplies conserved charges in involution along the classical trajectories.

Legendre transformations and auxiliary interactions are established tools in self-dual electrodynamics~\cite{Gaillard:1997rt,Ivanov:2003uj}. For the two-dimensional auxiliary field sigma model, the convention of Refs.~\cite{Ebert:2024zwv,Baglioni:2025tsc} uses the interaction $E(\nu)$ with $\nu=\sqrt{\Tr(v_+^2)\Tr(v_-^2)}$ and Lie-algebra-valued auxiliary fields $v_\pm$. On a patch with $E_{\nu\nu}\ne0$ and $|\mu|<1$, its dictionary is
\begin{equation}
\mu=E_\nu,\qquad H=E-\nu E_\nu=-\Omega(y),\qquad
 y=\frac{1+\mu}{1-\mu}.
\label{AFSMdictionary}
\end{equation}
Thus $\Omega$ corresponds to the transformed interaction, rather than to $E$ itself. For generalized Born--Infeld, $r=(\mu-t)/(1-\mu t)$ with $t=\tanh(\gamma/2)$; only at $\gamma=0$ does $r=\mu$. The linear ModMax interaction $E=t\nu$ is understood as the constrained limit $\mu=t$, consistently fixing $y=e^\gamma$~\cite{Bielli:2024ach,Baglioni:2025tsc}.

\section{Discussion}
For a prescribed one-scale CH family, a scalar inversion determines the autonomous classical generator and its single-valued physical domain; a regular auxiliary potential gives the same profile. This construction specializes known local stress-tensor inversions~\cite{Ferko:2023wyi} and gives exact profiles beyond finite-order matching~\cite{Babaei-Aghbolagh:2026vkm}. Equation~\eqref{profilestress} connects the profile directly to the relation between local energy density and pressure. Its concrete output is a deformation law, including its branch domain, for the chosen trajectory.

The pressure comparison is a direct consequence of the known Born--Infeld and logarithmic CH functions~\cite{Russo:2024llm}: their first irrelevant interactions agree, yet their pressures are strictly ordered at every positive energy density and have different strong-field limits. A pressure ceiling in one model and unbounded growth in the other survive the shared requirements of causality and classical integrability. This compares different full deformation laws with the same leading term; it asserts no ambiguity in a specified full $T\bar T$ flow. The shared CH structure also connects these sigma models to four-dimensional self-dual electrodynamics~\cite{Babaei-Aghbolagh:2025cni,Babaei-Aghbolagh:2025uoz,Baglioni:2025tsc}, subject to the respective stress-tensor identities and coupling normalizations.

The branch restrictions are essential to this classical result. Causality makes the implicit CH constraint regular, but does not guarantee global invertibility of $r(\zeta)$. Turning points require separate branches, and interchanging finite coupling changes requires both paths to remain on the same smooth physical branch of the common family. The nonpolynomial generators introduce a further question at the quantum level: how should stress-tensor composites such as $\sqrt{XY}$ be defined and renormalized? Whether generic reconstructed flows preserve quantum integrability remains open~\cite{Bielli:2026ggs}. The exact classical profiles provide specified deformation laws on which to pose that question.

\begin{acknowledgments}

We are grateful to Christian Ferko,  Jorge G. Russo, Dmitri Sorokin, and Roberto Tateo for their interest in this work and the fruitful discussions that followed. The work of H.B.-A. was conducted as part of the PostDoc Program on {\it Exploring TT-bar Deformations: Quantum Field Theory and Applications,} sponsored by Ningbo University, and this research was partly supported by NSFC Grants No.~12475053, 12235016 and 12588101.
\end{acknowledgments}

\bibliographystyle{apsrev4-2}
\bibliography{Ref_2D}

\end{document}